\documentclass{aa}
\usepackage[dvips]{graphicx}
\usepackage{txfonts}
\usepackage{natbib}
\bibpunct{(}{)}{;}{a}{}{,} 
\usepackage{subfigure}
\begin{document}

\authorrunning{J. P. Byrne et al.}

\titlerunning{The kinematics of CMEs using multiscale methods}

   \title{The kinematics of coronal mass ejections using multiscale methods}


   \author{J. P. Byrne\inst{1} \and P. T. Gallagher\inst{1} \and R. T. J. McAteer\inst{1} \and C. A. Young\inst{2}
          }

   \offprints{J. P. Byrne}

   \institute{Astrophysics Research Group, School of Physics, Trinity College Dublin,
			Dublin 2, Ireland.\\
              \email{jbyrne6@gmail.com}
        	\and
	ADNET Systems Inc., NASA Goddard Space Flight Center, 
           		Greenbelt, MD 20771, USA.\\
             }

   \date{Received 19 March 2008; accepted 16 December 2008}

 
  \abstract
  {}
   {The diffuse morphology and transient nature of coronal mass ejections (CMEs) make them difficult to identify and track using traditional image processing techniques. We apply multiscale methods to enhance the visibility of the faint CME front. This enables an ellipse characterisation to objectively study the changing morphology and kinematics of a sample of events imaged by the Large Angle Spectrometric Coronagraph (LASCO) onboard the Solar and Heliospheric Observatory (SOHO) and the Sun Earth Connection Coronal and Heliospheric Investigation (SECCHI) onboard the Solar Terrestrial Relations Observatory (STEREO). The accuracy of these methods allows us to test the CMEs for non-constant acceleration and expansion.}
   {We exploit the multiscale nature of CMEs to extract structure with a multiscale decomposition, akin to a Canny edge detector. Spatio-temporal filtering highlights the CME front as it propagates in time. We apply an ellipse parameterisation of the front to extract the kinematics (height, velocity, acceleration) and changing morphology (width, orientation).}
  {The kinematic evolution of the CMEs discussed in this paper have been shown to differ from existing catalogues. These catalogues are based upon running-difference techniques that can lead to over-estimating CME heights. Our resulting kinematic curves are not well-fitted with the constant acceleration model. It is shown that some events have high acceleration below $\sim$5~R$_{\sun}$. Furthermore, we find that the CME angular widths measured by these catalogues are over-estimated, and indeed for some events our analysis shows non-constant CME expansion across the plane-of-sky.}
   {}


   \maketitle
%

\section{Introduction}

Coronal mass ejections (CMEs) are large-scale eruptions of plasma and magnetic field with energies up to, and beyond, $10^{32}$~ergs.  The mass of particles expelled in a CME can amount to $10^{15}$~g, typically traveling from the Sun at velocities of hundreds up to several thousand kilometers per second \citep{Gosling76, Hundhausen94, Gopal01, Gallagher03}. They can lead to significant geomagnetic disturbances on Earth and may have negative impacts upon space-borne instruments susceptible to high levels of radiation when outside the protection of Earth's magnetosphere. While they have been investigated by remote sensing and in situ measurements for more than 30 years, their kinematic and morphological evolution through the corona and interplanetary space is still not completely understood.
\par
It is well known that CMEs are associated with filament eruptions and solar flares \citep{Zhang02, MoonChae04} but the driver mechanism remains elusive. Several theoretical models are described by \citet{Krall01} within the context of the magnetic flux-rope model of \citet{Chen96}. These include flux injection, magnetic twisting, magnetic energy release and hot plasma injection. The three-dimensional magnetic flux-rope model, in which a converging flow toward the neutral line results in reconnection beneath the flux-rope \citep{Forbes95, Amari03, Lin07}, assumes that the kinematics of an erupting flux-rope can be described using a force-balance equation, which includes gas pressure, gravity and the Lorentz force \citep{KrallChen05, Chen06, Manchester07}. An alternative to this is the magnetic break-out model in which the CME eruption is triggered by reconnection between the overlying field and a neighbouring flux system \citep{Antiochos99, Lynch04, MacNeice04}. The increased rate of outward expansion drives a faster rate of breakout reconnection yielding the positive feedback required for an explosive eruption. The models are dependent on geometrical properties of the CME, such as its width and radius, and they are designed to give an indication of the processes that drive CME kinematics. Thus it is important to develop methods of localising the CME front and characterising it in the observed data with high accuracy for model comparisons.
\par
To date, most CME kinematics are derived from running-difference images whereby each image is subtracted from the next in order to highlight regions of changing intensity. Once the CME is identified, either manually through point-and-click estimates, or by a form of image thresholding and segmentation, height-time plots are produced and the velocity and acceleration of each event determined. CMEs are often thought of as being either gradual or impulsive depending on both their eruption mechanism (e.g. prominence lift off or flare driven) and their speed \citep{Sheeley99, MoonChoe02}. The faster events tend to have an average negative acceleration attributed to aerodynamic drag of the solar wind \citep{Cargill04, Vrsnak04}. It was also shown by \citet{Gallagher03} and \citet{Zhang05} that CMEs may undergo an early impulsive acceleration phase below $\sim$2~R$_{\sun}$.
\par
Current methods of CME detection have their limitations, mostly since these diffuse objects have been difficult to identify using traditional image processing techniques. These difficulties arise from the transient nature of the CMEs, the scattering effects and non-linear intensity profile of the surrounding corona, and the interference of cosmic rays and solar energetic particles (SEPs) appearing as noise on the coronagraph detectors. Observations made by the Solar and Heliospheric Observatory's (SOHO) Large Angle Spectrometric Coronagraph (LASCO) are compiled into a CME catalog at the Coordinated Data Analysis Workshop \citep[CDAW;][]{Yashiro04} which operates by tracking the CME in LASCO/C2 and C3 running difference images to produce height-time plots of each event. It is a wholly manual procedure and is subject to user bias in interpreting the data. The Computer Aided CME Tracking routine \citep[CACTus;][]{Robbrecht04} is based upon LASCO/C2 and C3 running difference images. The images are unwrapped into polar coordinates and angular slices are stacked together in a time-height plot. CMEs thus appear as ridges in these plots, detected by a Hough Transform. The nature of this detection constrains the CMEs to have constant velocity and zero acceleration. The Solar Eruptive Event Detection Systems \citep[SEEDS;][]{Olmedo08} is an automatic detection based on LASCO/C2 running difference images unwrapped into polar coordinates. The algorithm uses a form of threshold segmentation to approximate the shape of the CME leading edge, and automatically determines the height, velocity and acceleration profiles.
\par
In this work we apply a new multiscale method of analysing CMEs. The use of multiscale methods in astrophysics have proven effective at denoising spectra and images \citep{Murtagh95, Fligge97}, analysing solar active region evolution \citep{Hewett08}, and enhancing solar coronal images \citep{Stenborg03, Stenborg08}. Multiscale analysis has the benefit of working on independent images without any need for differencing. A particular application of multiscale decompositions, implemented by \citet{Young08}, uses high and low pass filters convolved with the image data to exploit the multiscale nature of the CME. This highlights its intensity against the background corona as it propagates through the field-of-view, while neglecting small scale features (essentially denoising the data). \citet{Young08} also describe the use of nonmaxima suppression to trace the edges in the CME images, and they show the power of multiscale methods over previous edge detectors such as Roberts and Sobel. With these methods for defining the front of the CME we can characterise its morphology (width, orientation) and kinematics (position, velocity, acceleration) in coronagraph images. Following previous discussions on the errors in CME heights \citep[e.g.][]{Wen07}, multiscale methods allows us to determine the kinematics to a high degree of accuracy in order to confidently compare our results to theory. While these methods are not currently automated, they have great potential for such an implementation which the authors intend to pursue. 
\par
In Sect.~2 we discuss current image pre-processing methods from the coronagraphs onboard the SOHO \citep{Domingo97} and STEREO \citep{Kaiser08} spacecraft. We outline the implementation of multiscale methods and ellipse fitting to characterise the CME front. In Sect.~3 we present a sample of CME events analysed by these methods, and Sect.~4 is a discussion of these first results and conclusions.


\section{Observations \& Data Analysis}

\begin{figure}
\resizebox{\hsize}{!}{\includegraphics[clip=true, trim=0 0 0 460, width=\linewidth]{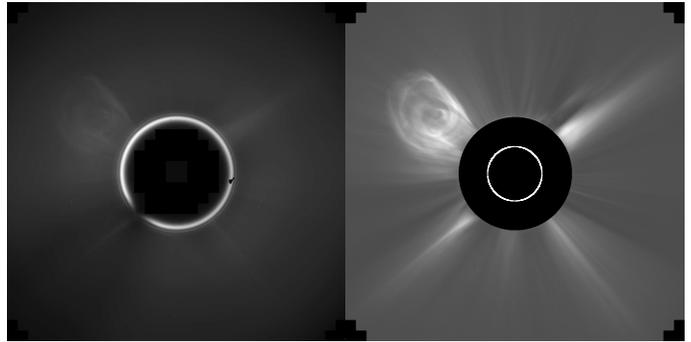}}
\caption{Raw (left) and pre-processed image (right) of a CME observed by LASCO on 2004 April 1. The pre-processing includes median filtering, background subtraction and occulter masking.}
\label{9811f1}
\end{figure}

In this paper a sample of CMEs observed by SOHO/LASCO and STEREO/SECCHI are studied, namely the gradual events of 2000 January 2, 2000 April 18, 2001 April 23, 2004 April 1, 2007 October 8 and 2007 November 16, and the impulsive events of 2000 April 23 and 2002 April 21.
\par
The LASCO/C2 and /C3 coronagraphs \citep{Brueckner95} onboard SOHO give a nested field-of-view extending from approximately 2.2 -- 30~R$_{\sun}$. The COR1/2 coronagraphs of the SECCHI suite \citep{Howard00} onboard STEREO image the corona from 1.4 -- 15~R$_{\sun}$. The image quality of the coronagraph observations can be diminished for many reasons, including instrumental effects (e.g. scattered light), noise from cosmic rays and SEPs, or data dropouts. In order to minimise these effects, we use the following standard preprocessing methods. Firstly the images are normalized with regard to exposure times in order to correct for temporal variations in the image statistics. Secondly a median filter is applied to remove pixel noise, replacing hot pixels with a median value of the surrounding pixel intensities. Finally we perform a background subtraction, obtained from the minimum of the daily median images across a time span of a month, and remove the occulting disc with a zero mask. These steps lead to a clear improvement in the image quality for CME study (Fig.~\ref{9811f1}), after which we apply our methods of multiscale analysis.
\par
In recent years the use of wavelets has been increasingly evident in image processing of solar structures \citep{Stenborg03, Stenborg08, Young08}. Here we illustrate briefly the formalism of the 1D wavelet transform, and discuss how this is extended to 2D. (For a more detailed discussion on wavelet transforms the reader is directed to e.g. \citet{Burrus98} and \citet{Lin04}.) The remainder of the section outlines our methods of using the multiscale analysis to define the CME front for characterisation.

\begin{figure}
\resizebox{\hsize}{!}{\includegraphics[clip=true, trim=0 100 0 100, width=\linewidth]{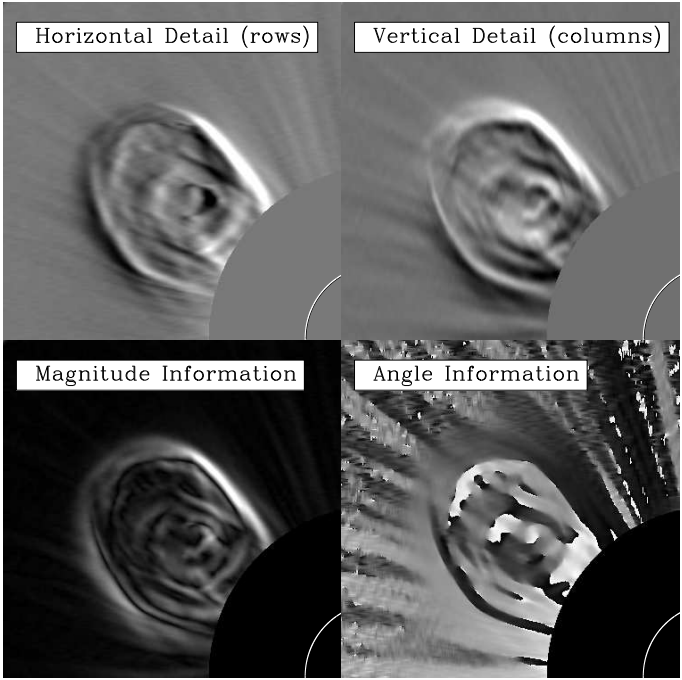}}
\caption{Top left, the horizontal, and top right, the vertical coefficients from the high-pass filtering at scale 3. Bottom left, the corresponding magnitude (edge strength) and bottom right, the angle information (0 -- 360$\degr$) taken from the gradient space, for a CME observed in LASCO/C2 on 2004 April 1.}
\label{9811f2}
\end{figure}

\subsection{Multiscale Filtering}

The continuous wavelet transform (CWT) of a signal $f(x)\in L^2(\mathbb{R})$ with respect to the mother wavelet $\psi(x) \in L^2(\mathbb{R})$ is defined by:
\begin{equation}
W(a,b)\;=\; \int_{-\infty}^\infty f(x) \psi_{a,b}(x)dx
\end{equation}
where $\psi_{a,b}(x)$ is a spatially localised function given by:
\begin{equation}
\psi_{a,b}(x)\;=\;\frac{1}{\sqrt{a}}\psi(\frac{x-b}{a})
\end{equation}
and $a,b \in L^2(\mathbb{R})$ correspond to the scaling (dilation) and shifting (translation) of $\psi(x)$. Note that the CWT spectrum obtained has no restriction as to how many scales are used, nor of the spacing between the scales. 

\begin{figure}
\centering
\subfigure[00:40 UT]
{
	\label{9811f3a}
	\includegraphics[scale=0.27, trim=105 120 0 120]{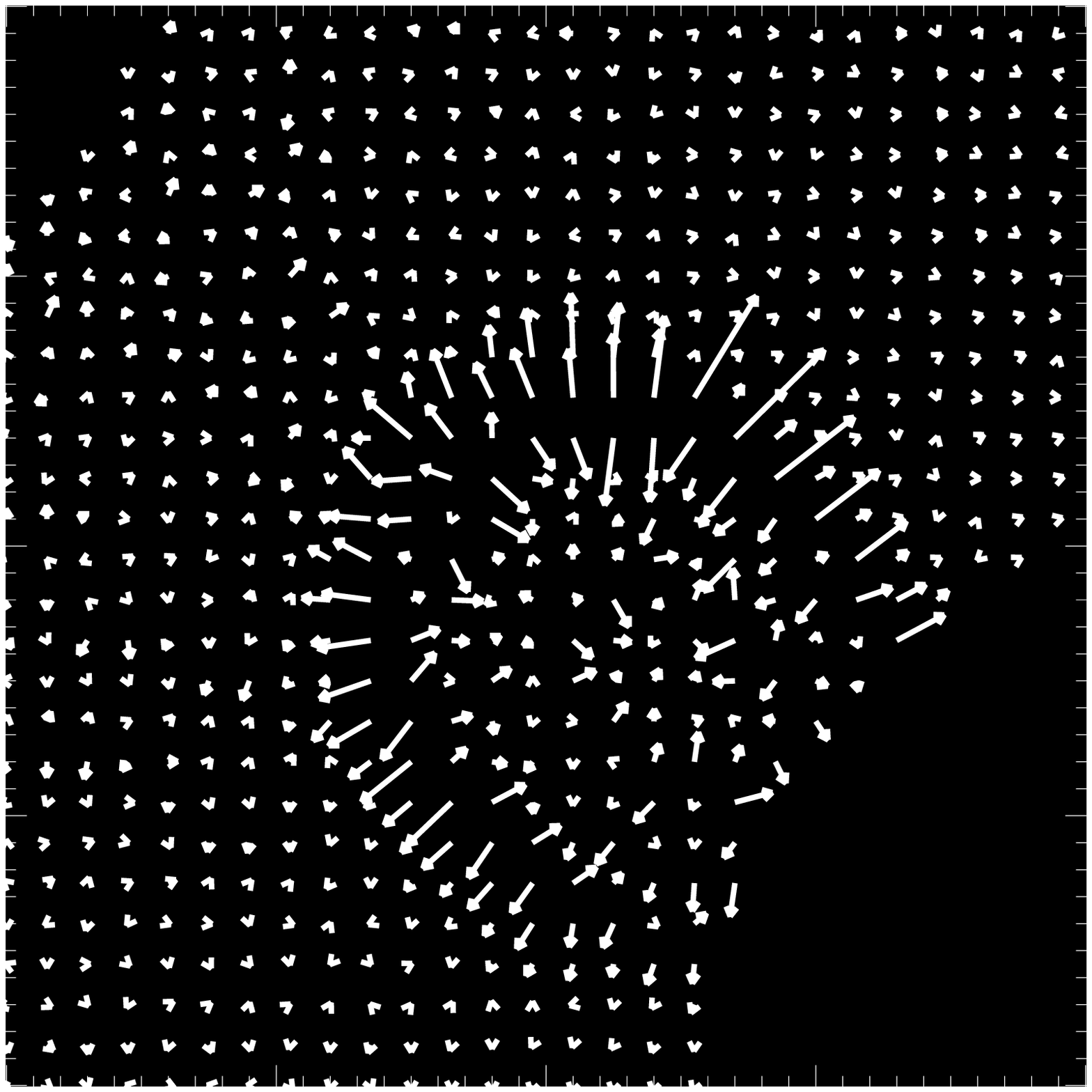}
		
}
\vspace{0cm}
\subfigure[01:00 UT]
{
	\label{9811f3b}
	\includegraphics[scale=0.27, trim=55 120 80 120]{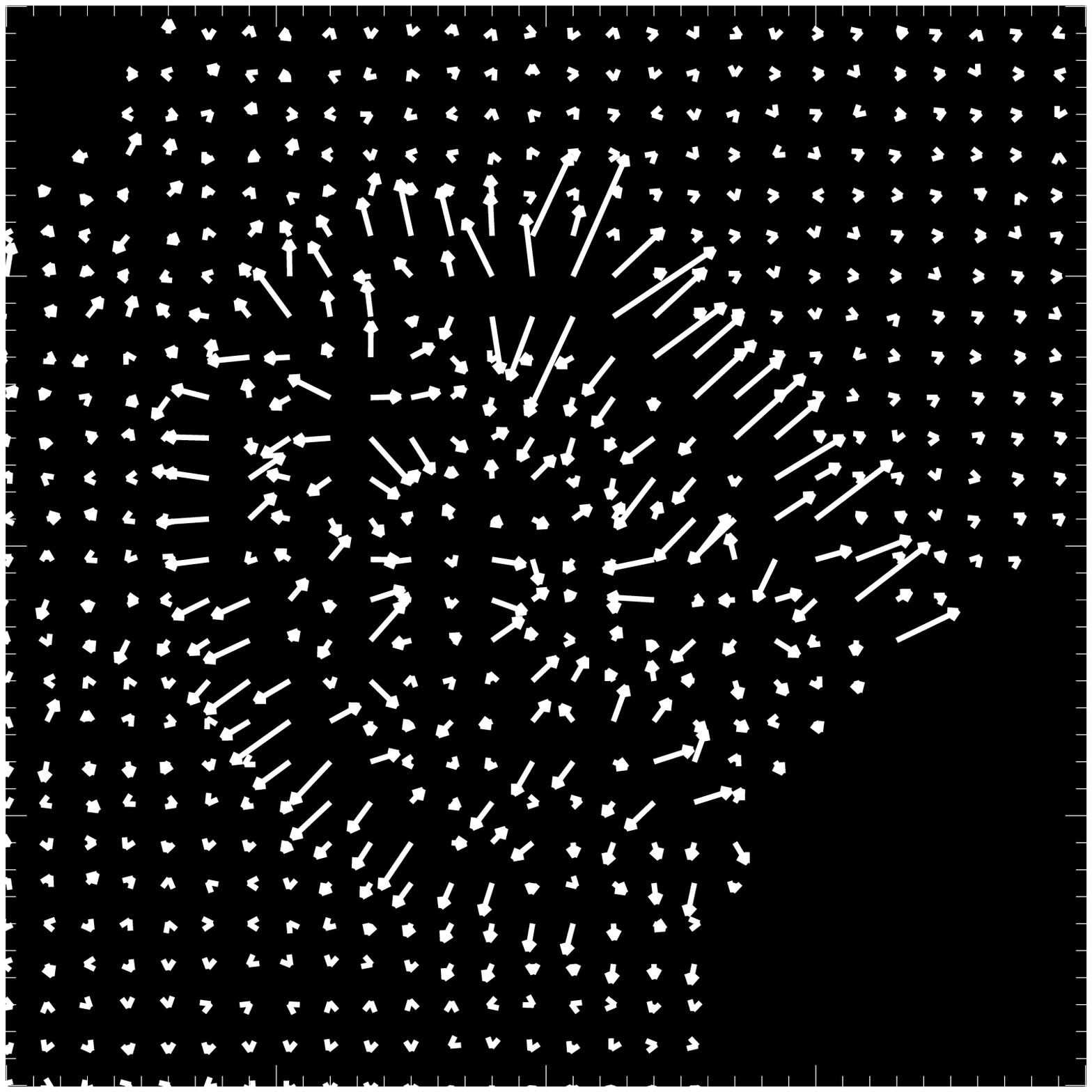}
}
\caption{The vectors plotted represent the magnitude and angle determined from the gradient space of the high-pass filtering at scale 3. The CME of 2004 April 1 shown here is highlighted very effectively by this method.}
\label{9811f3}
\end{figure}

On the other hand, the discrete wavelet transform (DWT) consists of a scaling function $\phi (x)$ and corresponding wavelet $\psi (x)$ with finite support $[0,l]$ ($l$ being a positive number) given by:
\begin{equation}
\phi (x) \;=\; \sqrt{2} \sum_{s=0}^l h_s \phi (2x-s) 
\end{equation}
\begin{equation}
\psi (x) \;=\; \sqrt{2} \sum_{s=0}^l g_s \phi (2x-s) 
\end{equation}
where $h$ and $g$ are constants called the low-pass and high-pass filter coefficients respectively. Unlike the CWT, the DWT decomposes a signal into a set of wavelet components with dyadic scaling, i.e. the scale increases in powers of 2 (2$^1$, 2$^2$, etc). In this way, the computational time required to perform the decomposition is drastically reduced.

We explore a method of multiscale decomposition by means of the DWT in 2D through the use of low and high pass filters using a discrete approximation of a Gaussian $\theta$ and its derivative $\psi$ respectively. Since $\theta(x,y)$ is separable we can write:
\begin{equation}
\psi_{x}(x,y)\;=\;\frac{\partial \theta(x)}{\partial x}\theta(y)
\end{equation}
\begin{equation}
\psi_{y}(x,y)\;=\;\theta(x)\frac{\partial \theta(y)}{\partial y}.
\end{equation}
Successive convolutions of an image with the filters produces the scales of decomposition, with the high-pass filtering providing the wavelet transform of image $I(x,y)$ in each direction:
\begin{equation}
W_{x}I\;=\;W_{x}I(x,y)\;=\;\psi_{x}(x,y)*I(x,y)
\end{equation}
\begin{equation}
W_{y}I\;=\;W_{y}I(x,y)\;=\;\psi_{y}(x,y)*I(x,y).
\end{equation}
Akin to a Canny edge detector \citep{Young08}, these horizontal and vertical wavelet coefficients are combined to form the gradient space $\Gamma$ for each scale: 
\begin{equation}
\Gamma (x,y)\; = \;\left [W_{x}I,~W_{y}I \right].
\end{equation}
The gradient information has an angular component $\alpha$ and a magnitude (edge strength) $M$:
\begin{equation}
\alpha(x,y) \; = \; tan^{-1}\left ( W_{y}I~/~W_{x}I \right )
\end{equation}
\begin{equation}
M(x,y) \; = \; \sqrt{ ( W_{x}I ) ^{2} + ( W_{y}I  ) ^{2} }.
\end{equation}
The resultant horizontal and vertical detail coefficients, and the magnitude and angular information are illustrated in Fig.~\ref{9811f2}.

\par
Overlaying a mesh of vector arrows on the data shows how the magnitude and angular information illustrate the progression of the CME. Each vector is rooted on a pixel in the gradient space, and has a length corresponding to the magnitude $M$ with an angle from the normal $\alpha$ (Fig.~\ref{9811f3}). By utilising all the information gained from the convolutions on the data array it becomes possible to threshold out the CME with a view to characterising its progression through space.

The magnitude information was found to have the highest signal-to-noise ratio at the third scale of the decomposition. This scale is very effective at smoothing unwanted artefacts such as cosmic rays which the median filter may have missed. The angular component $\alpha$ of the gradient specifies a direction which points across the greatest intensity change in the data (an edge). A threshold is specified with regard to this gradient direction in order to chain pixels along maxima, highlighting the edges in the image. A spatio-temporal threshold is implemented with regard to changes of magnitude in order to highlight moving features. It works by creating a specific detection mask which is used to pull out the edges along the CME front. In cases of faint CMEs or strong streamer deflections the filter is presently limited by working on only one scale, meaning current analysis involves a user removing/including certain edges that the algorithm has mistakenly retained/discarded. Future work on more than one scale may help alleviate this issue, since the additive angular distribution along a CME front is significantly larger than along a streamer when considered through multiple scales. Including this caveat will better enable the algorithm to automatically determine the CME edges, leading to a near real-time characterisation and kinematic analysis.

\subsection{Characterising the CME Front}

Using a model such as an ellipse to characterise the CME front across a sequence of images, has the benefit of providing the kinematics and morphology of a moving and/or expanding structure. The ellipse's multiple parameters, namely its changeable axes lengths and tilts, is adequate for approximating the varying curved structures of CMEs. \citet{Chen97} suggest an ellipse to be the two-dimensional projection of a flux rope, and \citet{KrallStCyr06} use ellipses to parameterise CMEs and explore their geometrical properties. We fit ellipses to the points determined to be along the CME front by considering a radial fan from Sun centre across the defined edges. This means there are more points along the front than on the flanks of the CME for inclusion in the fit, and the edges never double back on themselves in cases where the CME's internal structure might be otherwise included. A kinematic analysis provides height, velocity and acceleration profiles; while the ellipse's changing morphology provides the inclination angle and angular width. Measuring these properties in the observed data is vitally important for accurate comparison with theoretical models. 

\par
\begin{figure}
\resizebox{\hsize}{!}{\includegraphics[clip=true, trim=10 230 10 230, width=\linewidth]{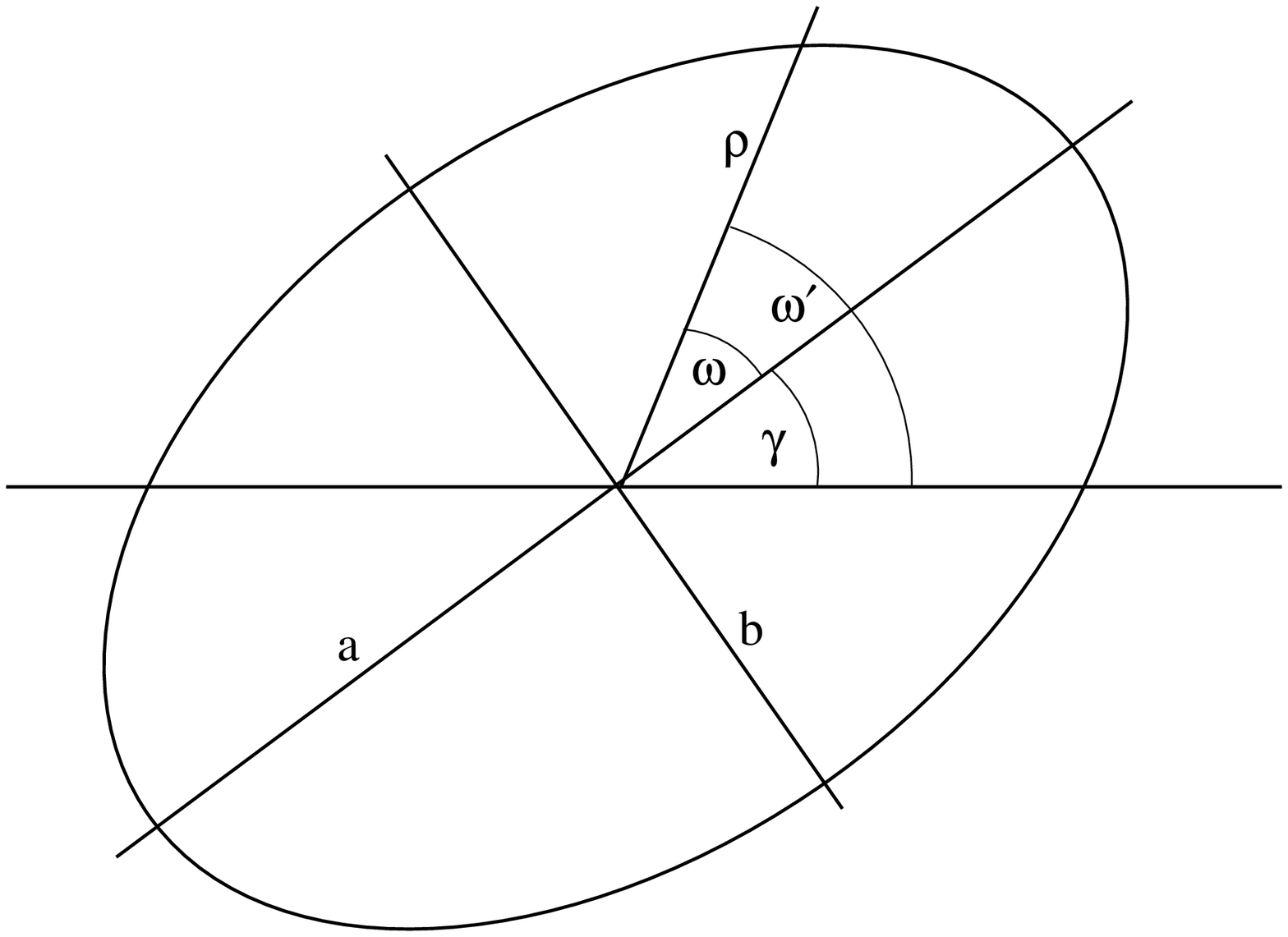}}
\caption{Ellipse inclined at angle $\gamma$, with semimajor axis $a$, semiminor axis $b$, and radial line $\rho$ inclined at angle $\omega$ to the semimajor axis.}
\label{9811f4}
\end{figure}

The implementation of the ellipse fitting routine is based upon an initial guess of ellipse centre as the average of the points specified along the front. The ellipse equation (in polar coordinates) is defined as:
\begin{equation}
\frac{\rho^2\cos^2\omega}{a^2}+\frac{\rho^2\sin^2\omega}{b^2}\;=\;1
\end{equation}
where $a$ and $b$ are the lengths of the semimajor and semiminor axes respectively, so allowing for an inclination angle \begin{math}\gamma\end{math} on the ellipse gives:
\begin{equation}
\rho^2\;=\;\frac{a^2b^2}{(\frac{a^2+b^2}{2})-(\frac{a^2-b^2}{2})\cos(2\omega'-2\gamma)}
\end{equation}
where \begin{math}\omega'=\omega+\gamma\end{math}, as illustrated in Fig.~\ref{9811f4}.
This gives a first approximation which can then be used to iteratively float the ellipse parameters until a Levenberg-Marquardt least-squares minimisation is reached.

\par
CME height measurements are taken as the height of the furthest point on the ellipse from Sun centre. The angular width is taken as the cone angle of the ellipse from Sun centre, and the tilt of the ellipse is given by the calculated angle $\gamma$. Note that in cases where the code produces an extremely large and oblate ellipse with one apex approximating the CME front, the width and tilt information is deemed redundant. Hence the resulting analysis of some events can have less data points included in the width and tilt plots than in the height-time plots.

\subsection{Error Analysis \& Model Fitting}
The front of the CME is determined through the multiscale decomposition and consequent rendering of a gradient magnitude space. At scale 3 of the decomposition the smoothing filter is $2^3$ pixels wide, which we use as our error estimate in edge position. This error is input to the ellipse fitting algorithm for weighting the ellipse parameters, and a final error output is produced for each ellipse fit. In the case of a fading leading edge the reduced amount of points along the front will increase the error on our analysis. The final errors are displayed in the height-time plots of the CMEs, and are used in the velocity and acceleration calculations. The derivative is a 3-point Lagrangian interpolation, so there is an enhancement of error at the edges of the data sets.
\par
The errors on the heights are used to constrain the best fit to a constant acceleration model of the form:
\begin{equation}
h(t)\;=\;a_0t^2+v_0t+h_0
\end{equation}
where $t$ is time and $a_0$, $v_0$ and $h_0$ are the acceleration, initial velocity and initial height respectively. This provides a linear fit to the derived velocity points and a constant fit to the acceleration. 
\par
An important point to note is the small time error (taken to be the image exposure time of the coronagraph data) since the analysis is performed upon the observed data frames individually. Previous methods of temporal-differencing would increase this time error. With these more accurate measurements we are better able to determine the velocity and acceleration errors, leading to improved constraints upon the data and providing greater confidence in comparing to theoretical models.

\section{Results}

This section outlines events which have been analyzed using our multiscale methods. We use data from the LASCO/C2 and C3, and SECCHI/COR1 and COR2 instruments, and preprocess the images as discussed in Sect.~2. The ellipse fitting algorithm applied to each event gives consistent heights of the CME front measured from Sun centre to the maximum height on the ellipse, and these lead to velocity and acceleration profiles of our events. The ellipse fitting also provides the angular widths and orientations, as shown below. The velocity, acceleration and angular width results of each method are highlighted in Tables 1, 2 and 3. In each instance we include the values from CACTus, CDAW and SEEDS, noting that CACTus lists a median speed of the CME; CDAW provide the speed at the final height and from the velocity profile we infer the speed at the initial height; and the SEEDS detection applies only to the LASCO/C2 field-of-view but doesn't currently provide a velocity range or profile. Note also that the CMEs of 2007 October 8 and 2007 November 16 are analysed in SECCHI images by CACTus and our multiscale methods (marked by asterisks in the Tables), while CDAW and SEEDS currently only provide LASCO analysis. It is clear that many of the CACTus, CDAW and SEEDS results lie outside the results and error ranges of our analysis.

\begin{table}[ht] 
\caption{Summary of CME velocities as measured by CACTus, CDAW, SEEDS and multiscale methods.} 
\centering      
\begin{tabular}{c c c c c}
\hline\hline                        
Date & CACTus & CDAW & SEEDS & Multiscale \\ [0.5ex] 
 & & km~s$^{-1}$ & & \\
\hline                    
2000 Jan 02 & 512 & 370 -- 794 & 396 & 396 -- 725  \\   
2000 Apr 18 & 463 & 410 -- 923 & 339 & 324 -- 1049  \\ 
2000 Apr 23 & 1041 & 1490 -- 898 & 595 & 1131 -- 1083  \\ 
2001 Apr 23 & 459 & 540 -- 519 & 501 & 581 -- 466 \\ 
2002 Apr 21 & 1103 & 2400 -- 2388 & 702 & 2195 -- 2412 \\
2004 Apr 01 & 487 & 300 -- 613 & 319 & 415 -- 570 \\
2007 Oct 08 & 235* & 85 -- 331 & 103 & 71 -- 330* \\
2007 Nov 16 & 337* & 210 -- 437 & 154 & 131 -- 483* \\ [1ex]       
\hline     
\end{tabular} 
*analysis of SECCHI data rather than LASCO.
\label{table:vel}  
\end{table}

\begin{table}[ht] 
\caption{Summary of CME accelerations as measured by CACTus, CDAW, SEEDS and multiscale methods.} 
\centering      
\begin{tabular}{c c c c c}
\hline\hline
Date & CACTus & CDAW & SEEDS & Multiscale \\ [0.5ex] 
& & m~s$^{-2}$ & & \\
\hline                    
2000 Jan 02 & 0 & 21.3 & $-$5.8 & 14.7~$\pm$~3.6  \\   
2000 Apr 18 & 0 & 23.1 & 17.5 & 32.3~$\pm$~3.5  \\ 
2000 Apr 23 & 0 & $-$48.5 & $-$8.9 & $-$4.8~$\pm$~20.6  \\ 
2001 Apr 23 & 0 & $-$0.7 & $-$1.4 & $-$4.8~$\pm$~4.1 \\ 
2002 Apr 21 & 0 & $-$1.4 & 33.5 & 32.5~$\pm$~26.6 \\
2004 Apr 01 & 0 & 7.1 & 12.9 & 4.4~$\pm$~2.0 \\
2007 Oct 08 & 0* & 3.4 & 2.4 & 5.7~$\pm$~0.9* \\
2007 Nov 16 & 0* & 4.9 & 11.0 & 13.7~$\pm$~1.7* \\ [1ex]       
\hline     
\end{tabular}
*analysis of SECCHI data rather than LASCO.
\label{table:accel}  
\end{table}

\begin{table}[ht] 
\caption{Summary of CME angular widths as measured by CACTus, CDAW, SEEDS and multiscale methods.}  
\centering      
\begin{tabular}{c c c c c}
\hline\hline
Date & CACTus & CDAW & SEEDS & Multiscale \\ [0.5ex] 
 & & degrees & & \\ 
\hline                    
2000 Jan 02 & 160 & 107 & 96 & 50 -- 95  \\   
2000 Apr 18 & 106 & 105 & 108 & 68 -- 110  \\ 
2000 Apr 23 & 352 & 360 & 130 & 96 -- 130  \\ 
2001 Apr 23 & 124 & 91 & 74 & 55 -- 60 \\ 
2002 Apr 21 & 352 & 360 & 186 & 53 -- 65 \\
2004 Apr 01 & 66 & 79 & 58 & 44 -- 38 \\
2007 Oct 08 & 52* & 82 & 59 & 23 -- 60* \\
2007 Nov 16 & 68* & 78 & 54 & 40 -- 55* \\ [1ex]       
\hline     
\end{tabular}
*analysis of SECCHI data rather than LASCO.
\label{table:aw}  
\end{table}

\begin{figure*}
\resizebox{\hsize}{!}{\includegraphics[clip=true, trim=0 20 0 0]{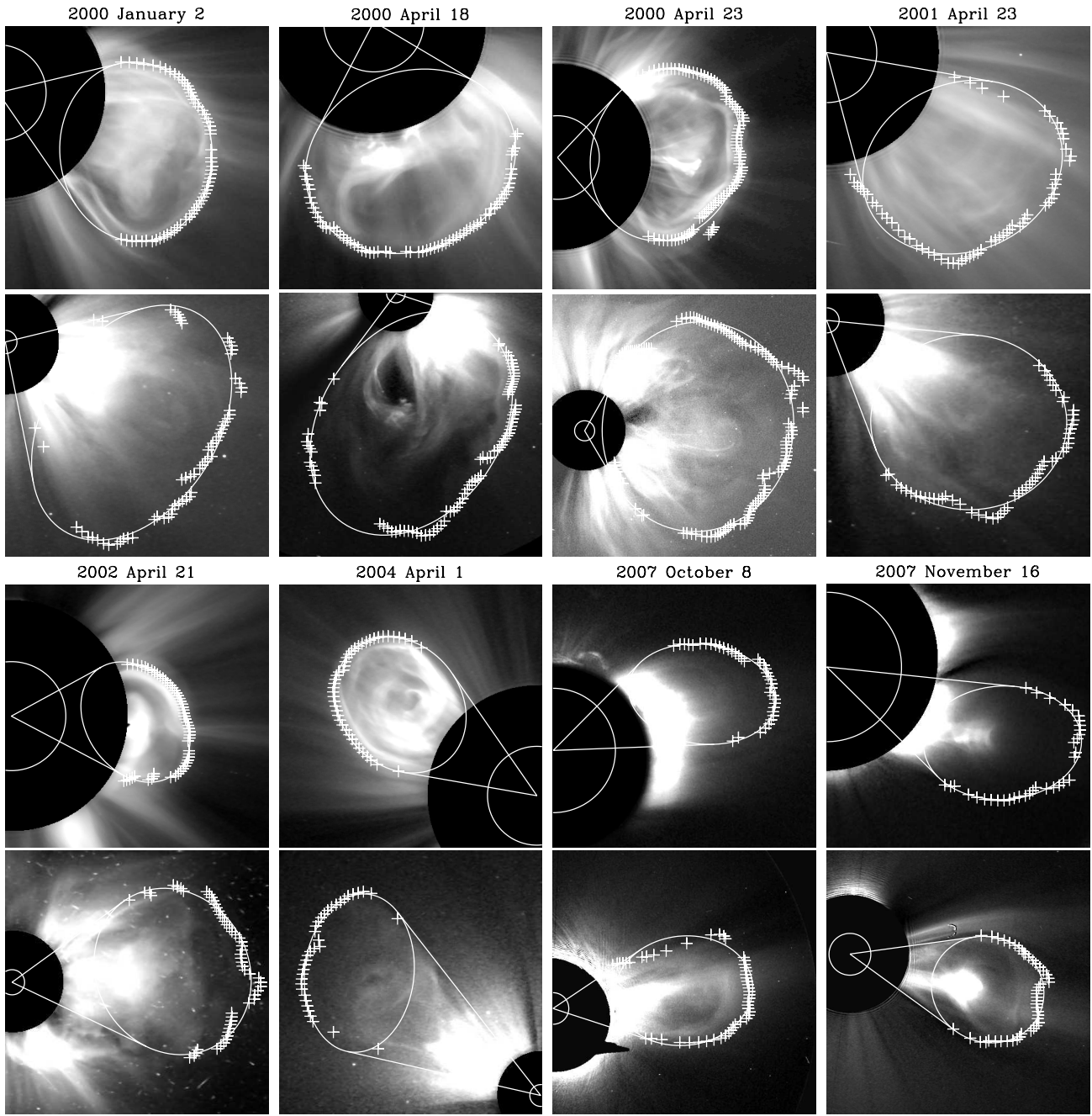}}
\caption{A sample of ellipse fits to the multiscale edge detection of the events studied. For each event the upper and lower image show LASCO/C2 and /C3 (except for the 2007 events which show SECCHI/COR1 and /COR2).}
\label{9811f5}
\end{figure*}

\subsection{Arcade Eruption: 2000 January 2}
This CME was first observed in the south-west at 06:06 UT on 2000 January 2 and appears to be a far-side event associated with an arcade eruption consisting of one or more bright loops. 
\par
\begin{figure}
\resizebox{\hsize}{!}{\includegraphics[clip=true, trim=0 100 0 10, width=\linewidth]{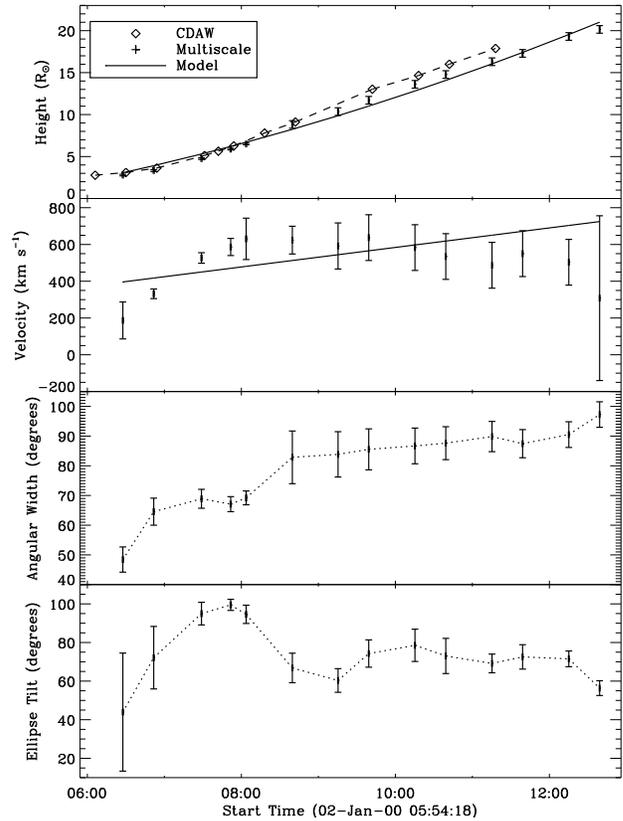}}
\caption{Kinematic and morphological profiles for the ellipse fit to the multiscale edge detection of the 2000 January 2 CME observed by LASCO/C2 and C3. The plots from top to bottom are height, velocity angular width, and ellipse tilt. The CDAW heights are over-plotted with a dashed line. The height and velocity fits are based upon the constant acceleration model.}
\label{9811f6}
\end{figure}

The height-time plot has a trend not unlike that of CDAW (overplotted in top Fig.~\ref{9811f6} with a dashed line). However the offset of the CDAW heights - which puts them outside our error bounds - may be due to how the difference images are scaled for display. This is a problem multiscale methods avoid. From Fig.~\ref{9811f6}, the velocity-fit was found to be increasing from 396 to 725~km~s$^{-1}$, giving an acceleration of 14.7~$\pm$~3.6~m~s$^{-2}$. The ellipse fit spans approximately 50 -- 70$\degr$ of the field-of-view in the inner portion of C2, and expands to over 95$\degr$ in C3. This expansion may simply be attributed to the inclusion of one or more loops in the ellipse fit as the arcade traverses the LASCO/C2 and C3 fields-of-view.  The orientation of the ellipse as a function of time is shown in bottom Fig.~\ref{9811f6}. It can be seen that the orientation angle of the CME increases to approximately 100$\degr$ before decreasing toward 60$\degr$. 
\par
The constant acceleration model is not a sufficient fit to the data in this event. The kinematics produced from the multiscale edge detection would be better fit with a non-linear velocity and a non-constant acceleration. This would show the CME to have a period of decreasing acceleration in the C2 field-of-view, leveling off to zero in C3 (if not decelerating further).


\subsection{Gradual/Expanding CME: 2000 April 18}
This CME was first observed off the south limb at 16:06 UT on 2000 April 18 and exhibits a flux rope type structure.
\par
\begin{figure}
\resizebox{\hsize}{!}{\includegraphics[clip=true, trim=0 100 0 10, width=\linewidth]{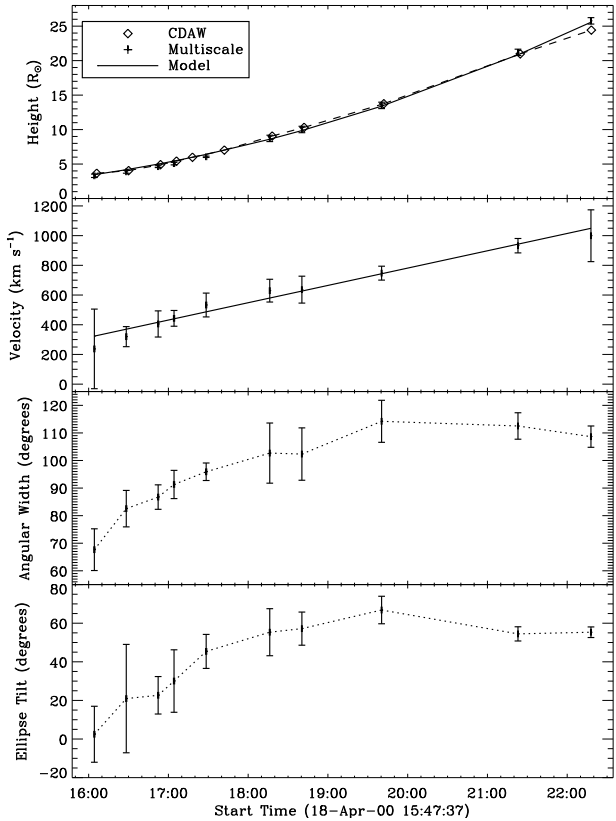}}
\caption{Same as Fig.~\ref{9811f6} but for the gradual CME of 2000 April 18.}
\label{9811f7}
\end{figure}

The height-time plot for this event has a trend similar to that of CDAW (overplotted in top Fig.~\ref{9811f7} with a dashed line). The velocity-fit was found to be linearly increasing from 324 to over 1\,000~km~s$^{-1}$, giving an acceleration of 32.3~$\pm$~3.5~m~s$^{-2}$. The ellipse fit spans from 68$\degr$ of the field-of-view in the inner portion of C2, to approximately 110$\degr$ in C3.  The orientation of the ellipse as a function of time is shown to increase from just above 0$\degr$ to over 60$\degr$ in Fig.~\ref{9811f7}. 
\par
This CME is fitted well with the constant acceleration model but shows an increasing angular width implying expansion across the field-of-view.

\subsection{Impulsive CME: 2000 April 23}
This impulsive CME was first observed in the west at 12:54 UT on 2000 April 23 and exhibits strong streamer deflection.  
\par
\begin{figure}
\resizebox{\hsize}{!}{\includegraphics[clip=true, trim=0 100 0 10, width=\linewidth]{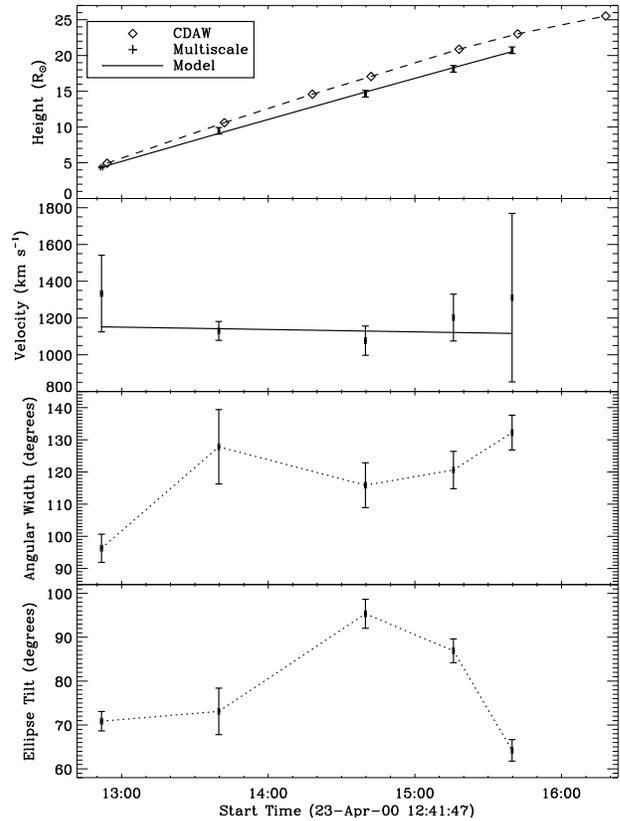}}
\caption{Same as Fig.~\ref{9811f6} but for the impulsive CME of 2000 April 23.}
\label{9811f8}
\end{figure}

The height-time plot derived using our methods has a trend which diverges from that of CDAW (overplotted in top Fig.~\ref{9811f8} with a dashed line). The velocity-fit was found to be linearly decreasing from 1\,131 to 1\,083~km~s$^{-1}$, giving a constant deceleration of $-$4.8~$\pm$~20.6~m~s$^{-2}$. The CME is present for one frame in C2 with an ellipse fit spanning 96$\degr$, increasing to approximately 120 -- 130$\degr$ in the C3 field-of-view, and the orientation of the ellipse as a function of time is shown to rise from 71$\degr$ to 95$\degr$ then fall to 64$\degr$ (see bottom Fig.~\ref{9811f8}). 
\par
This event is modeled satisfactorily with a constant deceleration. However, due to the impulsive nature of the CME there are only a few frames available for analysis, making it difficult to constrain the kinematics. 

\subsection{Faint CME: 2001 April 23}
This CME was first observed in the south-west at 12:39 UT on 2001 April 23 and exhibits some degree of streamer deflection. 
\par
\begin{figure}
\resizebox{\hsize}{!}{\includegraphics[clip=true, trim=0 100 0 10, width=\linewidth]{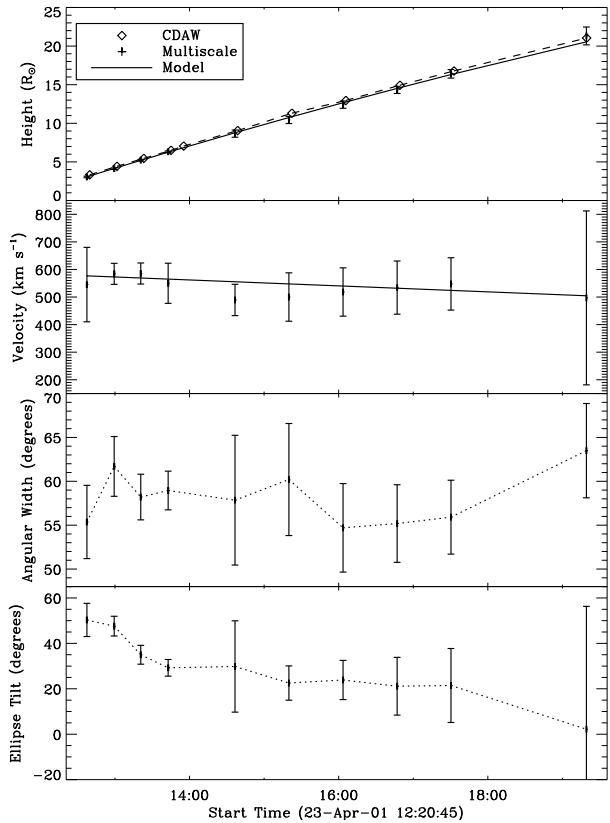}}
\caption{Same as Fig.~\ref{9811f6} but for the faint CME of 2001 April 23.}
\label{9811f9}
\end{figure}

The height-time plot has a similar trend to CDAW (overplotted in top Fig.~\ref{9811f9} with a dashed line). The velocity-fit was found to be linearly decreasing from 581 to 466~km~s$^{-1}$, giving a deceleration of $-$4.8~$\pm$~4.1~m~s$^{-2}$. The ellipse fit spans approximately 55 -- 60$\degr$ of the field-of-view throughout the event, and the orientation of the ellipse as a function of time is shown to decrease from approximately 50$\degr$ to almost 0$\degr$ (see Fig.~\ref{9811f9}).
\par
This CME is fitted well with the constant acceleration model.

\subsection{Fast CME: 2002 April 21}
This CME was first observed in the west from 01:27 UT on 2002 April 21. 
\par
\begin{figure}
\resizebox{\hsize}{!}{\includegraphics[clip=true, trim=0 100 0 10, width=\linewidth]{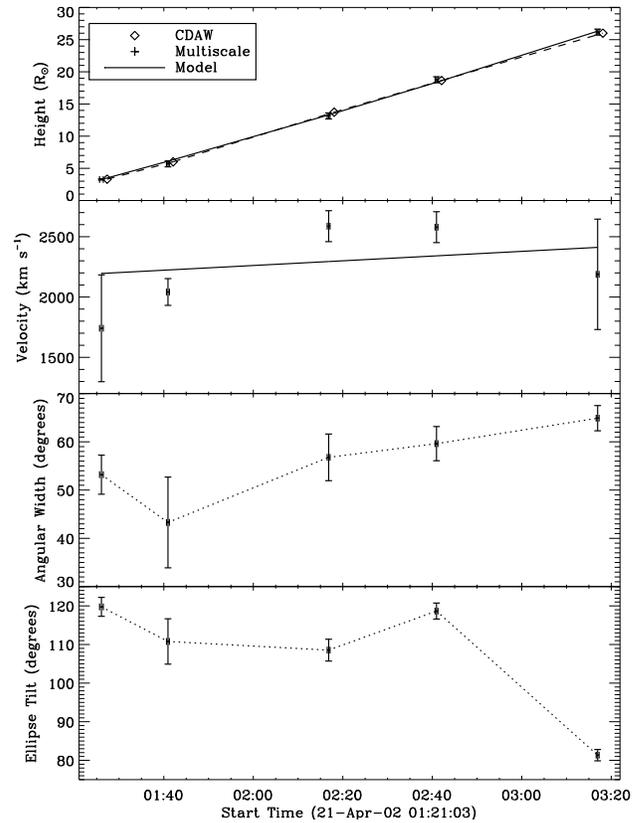}}
\caption{Same as Fig.~\ref{9811f6} but for the fast CME of 2002 April 21.}
\label{9811f10}
\end{figure}

The height-time plot follows a similar trend to that of CDAW (overplotted in top Fig.~\ref{9811f10} with a dashed line). The velocity-fit was found to be linearly increasing from 2\,195 to 2\,412~km~s$^{-1}$, giving a constant acceleration fit of 32.5~$\pm$~26.6~m~s$^{-2}$. The ellipse fit spans 53$\degr$ in C2, and shows an likely increasing trend to 65$\degr$ in C3.  The orientation of the ellipse as a function of time is shown to scatter about 115$\degr$ though it drops to approximately 81$\degr$ in the final C3 image.
\par
The kinematics of this event are not modeled satisfactorily by the constant acceleration model, since the fits do not lie within all error bars. The argument for a non-linear velocity profile, with a possible early decreasing acceleration, is justified for this event, although the instrument cadence limits the data set available for interpretation. The previous analysis of \citet{Gallagher03} resulted in a velocity of $\sim$2\,500~km~s$^{-1}$ past $\sim$3.4~R$_{\sun}$ which is consistent with our results past $\sim$6~R$_{\sun}$ in Fig.~\ref{9811f10}.

\subsection{Flux-Rope/Slow CME: 2004 April 1}
This CME was first observed in the north-east from approximately 23:00 UT on 2004 April 1, is in the field-of-view for over 9 hours, and exhibits a bright loop front, cavity and twisted core. 
\par
\begin{figure}
\resizebox{\hsize}{!}{\includegraphics[clip=true, trim=0 100 0 10, width=\linewidth]{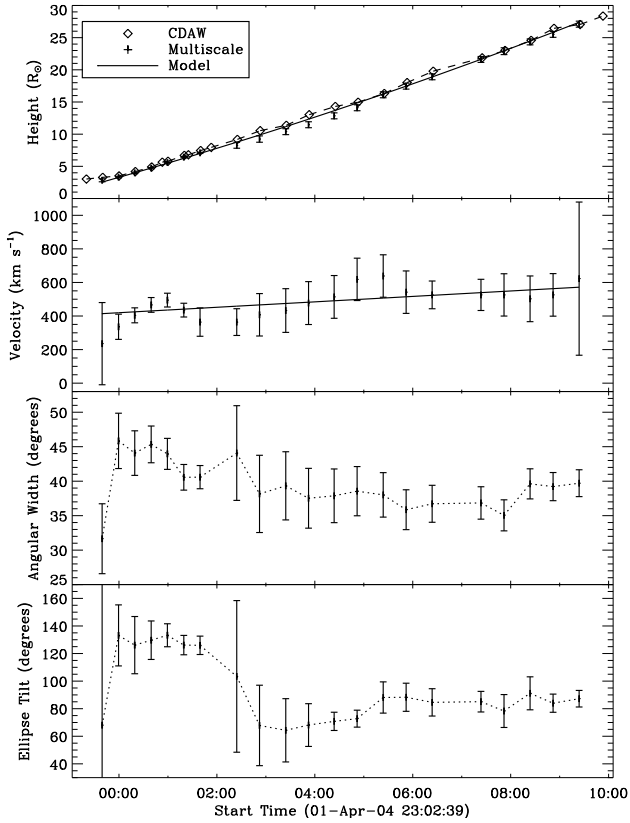}}
\caption{Same as Fig.~\ref{9811f6} but for the slow CME of 2004 April 1.}
\label{9811f11}
\end{figure}

The height-time plot follows a similar trend to that of CDAW (overplotted in top Fig.~\ref{9811f11} with a dashed line). The velocity-fit was found to be linearly increasing from 415 to 570~km~s$^{-1}$, giving an acceleration of 4.4~$\pm$~2.0~m~s$^{-2}$. Note also that the kinematics of this event exhibit non-linear structure clearly seen in the velocity and acceleration profiles. The ellipse fit spans approximately 44$\degr$ in C2, stepping down to approximately 38$\degr$ in C3.  The orientation of the ellipse as a function of time is shown to jump down from approximately 130$\degr$ in C2 to approximately 70--80$\degr$ in C3.
\par
This event shows unexpected structure in the velocity and acceleration profiles which indicates a complex eruption not satisfactorily modeled with constant acceleration.

\subsection{STEREO-B Event: 2007 October 8}
This CME was first observed in the west from approximately 12:00 UT on 2007 October 8, and is best viewed from the STEREO-B spacecraft. It is noted that the kinematics as measured by SOHO and STEREO will be different due to projection effects \citep{Vrsnak07}. It is intended that these effects be measured and corrected for in future work, especially as the spacecraft separations increase over time. On this date STEREO-B was at an angular separation of 16.5$\degr$ from Earth.
\par
\begin{figure}
\resizebox{\hsize}{!}{\includegraphics[clip=true, trim=0 100 0 10, width=\linewidth]{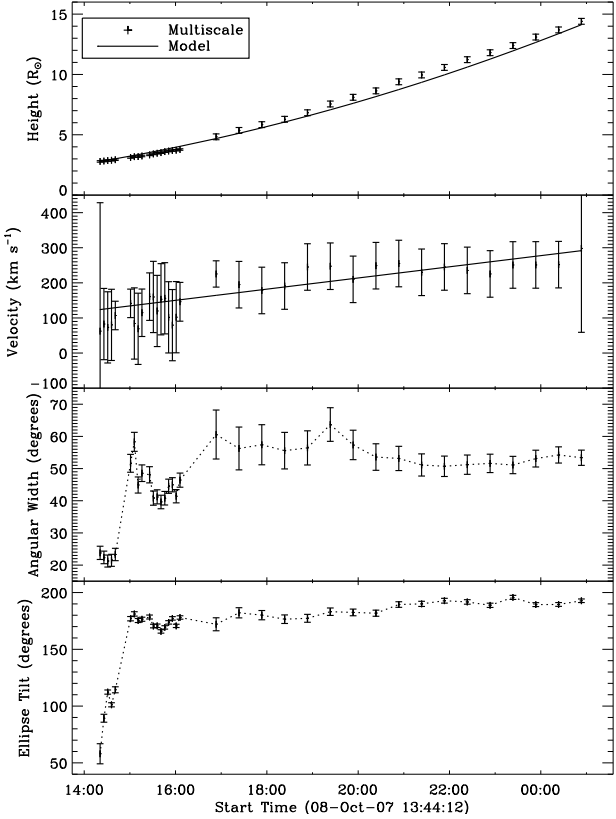}}
\caption{Same as Fig.~\ref{9811f6} but for the CME of 2007 October 8 observed by SECCHI/COR1 and COR2. For this reason CDAW is not overplotted.}
\label{9811f12}
\end{figure}

The height-time plot for this event is shown in Fig.~\ref{9811f12}. The velocity-fit was found to be linearly increasing from 71 to 330~km~s$^{-1}$, giving an acceleration of 5.7~$\pm$~0.9~m~s$^{-2}$. The ellipse fit in COR1 spans approximately 23$\degr$ stepping up to a scatter about 40 -- 50$\degr$ which rises slightly to 50 -- 60$\degr$ in COR2.  The orientation of the ellipse as a function of time is shown to increase from 55 -- 110$\degr$ then jumps to an approximately steady scatter about 180 -- 190$\degr$. 
\par
This CME is fitted well with the constant acceleration model.

\subsection{STEREO-A Event: 2007 November 16}
This CME was first observed in the west from approximately 08:26 UT on 2007 November 16, and is best viewed from the STEREO-A spacecraft. On this date STEREO-A was at an angular separation of 20.3$\degr$ from Earth.
\par
\begin{figure}
\resizebox{\hsize}{!}{\includegraphics[clip=true, trim=0 100 0 10, width=\linewidth]{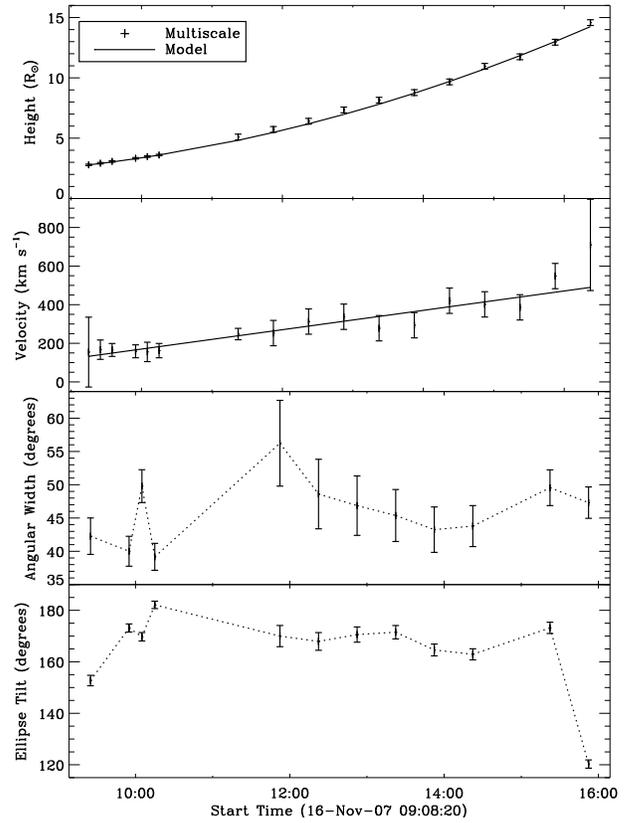}}
\caption{Same as Fig.~\ref{9811f12} but for the CME of 2007 November 16.}
\label{9811f13}
\end{figure}

The height-time plot for this event is shown in Fig.~\ref{9811f13}. The velocity-fit was found to be linearly increasing from 131 to 483~km~s$^{-1}$, giving an acceleration of 13.7~$\pm$~1.7~m~s$^{-2}$. The ellipse fit in COR1 spans approximately 40 -- 50$\degr$ stepping up slightly to a scatter about 45 -- 55$\degr$ in COR2.  The orientation of the ellipse as a function of time is shown to start at 153$\degr$ and end at 120$\degr$ with the mid points scattered about 170$\degr$. 
\par
This CME is fitted well with the constant acceleration model.

%

\section{Discussion \& Conclusions}

A variety of theoretical models have been proposed to describe CMEs, especially their early propagation phase. Observational studies, such as those outlined above, are necessary to determine CME characteristics. We argue that the results of previous methods are limited in this regard due mainly to large kinematic errors which fail to constrain a model, an artefact of CME detection based upon either running- (or fixed-) difference techniques or other operations. Current methods fit either a linear model to the height-time curve, implying constant velocity and zero acceleration (e.g. CACTus) or a second order polynomial, producing a linear velocity and constant acceleration (e.g. CDAW, SEEDS). The implementation of a multiscale decomposition provides a time error on the scale of seconds (the exposure time of the instrument) and a resulting height error on the order of a few pixels. The height-time error is used to determine the errors of the velocity and acceleration profiles of the CMEs. It was shown that for certain events the results of CACTus, CDAW and SEEDS can differ significantly from our methods, as illustrated in the Tables of Sect.~3.
\par
Our results clearly confirm that the constant acceleration model may not always be appropriate. The 2000 January 2 and 2002 April 21 CMEs are good examples of the possible non-linear velocity profile and consequent non-constant acceleration profile (see Fig.~\ref{9811f6} and Fig.~\ref{9811f10}). Indeed these events are shown to have a decreasing acceleration, possibly to zero or below, as the CMEs traverse the field-of-view. Simulations of the breakout model outlined in \citet{Lynch04} resulted in constant acceleration fits which do not agree with these observations. It may be further noted that the events of 2001 April 23 and 2004 April 1 show a possible decreasing acceleration phase early on, though within errors this cannot be certain (see bottom Fig.~\ref{9811f9} and Fig.~\ref{9811f11}). Furthermore, the structure seen in some events would indicate that the CME does not progress smoothly. The velocities of the 2004 April 1 CME in Fig.~\ref{9811f11} and the 2007 November 16 CME in Fig.~\ref{9811f13} show non-smooth profiles and may imply a form of bursty reconnection or other staggered energy release driving the CME. Other profiles such as Fig.~\ref{9811f6} and to a lesser extent Figs.~\ref{9811f9} and \ref{9811f10} may show a stepwise pattern, indicative of separate regimes of CME progression. None of the current CME models indicate a form of non-smooth progression, although the flux-rope model does describe an early acceleration regime giving a non-linear velocity to the eruption \citep[see Fig.~11.5 in][]{PriestForbesBook}.
\par
It may be concluded that the angular widths of the events are indicative of whether the CME expands radially or otherwise in the plane-of-sky. For the CMEs studied above, the observations of 2000 April 18, 2000 April 23 and 2002 April 21 show a super-radial expansion (see Fig.~\ref{9811f7}, Fig.~\ref{9811f8} and Fig.~\ref{9811f10}). These events also show high velocities, obtaining top speeds of up to 1\,000~km~s$^{-1}$, over 1\,100~km~s$^{-1}$ and 2\,500~km~s$^{-1}$ respectively, and may therefore indicate a link between the CME expansion and speed. Furthermore, it is suggested by \citet{KrallStCyr06} that the flux-rope model can account for different observed expansion rates due to the axial versus broadside view of the erupting flux system. 
\par
The observed morphology of the ellipse fits may be further interpreted through the tilt angles plotted in Sect.~3. In knowing the ellipse tilt and the direction of propagation of the CME it is possible to describe the curvature of the front. For the events above, the changing tilt and hence curvature is possibly significant for the 2000 April 18, 2004 April 1 and 2007 October 8 events (see bottom Fig.~\ref{9811f7}, Fig.~\ref{9811f11} and Fig.~\ref{9811f12}). The elliptical flux rope model of \citet{Krall05} was shown to have a changing orientation of the magnetic axis which results in a dynamic radius of curvature of the CME, possibly accounting for these observed ellipse tilts.
\par
The work outlined here is an initial indication that the zero and constant acceleration models in CME analysis are not an accurate representation of all events, and the over-estimated angular widths are not indicative of the true CME expansion. The ellipse characterisation has provided additional information on the system through its changing width and orientation which shall be further investigated with the analysis of many more events. This work will be further explored and developed with STEREO data whereby the combined view-points can give additional kinematic constraints and may lead to a correction for projection effects and possible three dimensional reconstructions.

\begin{acknowledgements}
      This work is supported by grants from Science Foundation Ireland's Research Frontiers Programme and NASA's Living With A Star Program. JMA is grateful to the STEREO/COR1 team at NASA/GSFC and is currently funded by a Marie Curie Fellowship. We would also like to thank the anonymous referee for their very helpful comments. SOHO is a project of international collaboration between ESA and NASA. The STEREO/SECCHI project is an international consortium of the Naval Research Laboratory (USA), Lockheed Martin Solar and Astrophysics Lab (USA), NASA Goddard Space Flight Center (USA), Rutherford Appleton Laboratory (UK), University of Birmingham (UK), Max-Planck-Institut f\"{u}r Sonnen-systemforschung (Germany), Centre Spatial de Liege (Belgium), Institut d'Optique Th\'{e}orique et Appliqu\'{e}e (France), and Institut d'Astrophysique Spatiale (France).
\end{acknowledgements}

\bibliographystyle{aa.bst}
\bibliography{9811Byrn}

\end{document}